\documentclass[aps,prl,twocolumn,showpacs,floatfix,nofootinbib,preprintnumbers,superscriptaddress,amsmath,amssymb]{revtex4-1}

\usepackage{graphicx}
\usepackage{epsfig}
\usepackage{bm}
\usepackage{color}
\usepackage{float}
\usepackage{dcolumn}
\usepackage{multirow}

\newcommand{\UNEDFZERO}{\textsc{unedf0}}
\newcommand{\UNEDFONE}{\textsc{unedf1}}
\newcommand{\HFBTHO}{\textsc{hfbtho}}

\newcommand{\nskin}{$r_{\rm skin}$}
\newcommand{\nskineq}{r_{\rm skin}}

\newcommand{\asym}{a_\text{sym}}

\begin{document}



\title{Neutron skin uncertainties of Skyrme energy density functionals}

\author{M. Kortelainen}
\affiliation{Department of Physics, P.O. Box 35 (YFL), FI-40014 University of Jyv\"askyl\"a, Finland}
\affiliation{Department of Physics and Astronomy, University of Tennessee, Knoxville, Tennessee 37996, USA}
\author{J. Erler}
\affiliation{Division Biophysics of Macromolecules, German Cancer Research Center (DKFZ), Im Neuenheimer Feld 580, D-69120 Heidelberg, Germany}
\author{W. Nazarewicz}
\affiliation{Department of Physics and Astronomy, University of Tennessee, Knoxville, Tennessee 37996, USA}
\affiliation{Physics Division, Oak Ridge National Laboratory,  Oak Ridge, Tennessee 37831, USA}
\affiliation{Faculty of Physics, University of Warsaw, ul. Ho\.za 69, 00-681 Warsaw, Poland}
\author{N. Birge}
\affiliation{Department of Physics and Astronomy, University of Tennessee, Knoxville, Tennessee 37996, USA}
\author{Y. Gao} 
\affiliation{Department of Physics, P.O. Box 35 (YFL), FI-40014 University of Jyv\"askyl\"a, Finland}
\author{E. Olsen}
\affiliation{Department of Physics and Astronomy, University of Tennessee, Knoxville, Tennessee 37996, USA}
%


\begin{abstract}
\begin{description}
\item[Background]
Neutron-skin thickness is an excellent indicator of isovector properties of atomic nuclei. As such, it correlates strongly with observables in finite nuclei that depend on neutron-to-proton imbalance and the nuclear symmetry energy  that characterizes the equation of state of neutron-rich matter. 
A rich worldwide experimental program involving studies with rare isotopes, parity violating electron scattering, and astronomical observations is devoted to pinning down the isovector sector of nuclear models.
\item[Purpose]
We assess the theoretical systematic and statistical uncertainties of neutron-skin thickness and relate them to the equation of state of nuclear matter, and in particular to nuclear symmetry energy parameters.
\item[Methods]
We use the nuclear superfluid Density Functional Theory
with several Skyrme energy density functionals and density dependent pairing. To evaluate statistical errors and their budget, we employ the statistical covariance technique.
\item[Results] 
We find that the errors on neutron skin increase with neutron excess.
Statistical errors due to uncertain coupling constants of the density functional  are found to be larger than systematic errors, the latter not exceeding 0.06\,fm in most neutron-rich nuclei across the nuclear landscape. The single major source of uncertainty is the poorly determined slope $L$ of the symmetry energy that parametrizes its density dependence.
\item[Conclusions]
To provide essential constraints on the symmetry energy of the  nuclear energy density functional, next-generation measurements of neutron skins are required to deliver precision better than 0.06\,fm. 
\end{description}
\end{abstract}


\maketitle


{\it Introduction} ---
The radioactive beam facilities of the  next generation  will enter the vast, currently unexplored territory of the nuclear landscape towards its limits \cite{(Erl12)}. This voyage is not going to be easy, especially on the neutron-rich side, but the scientific payoff is expected to be rich and multifaceted \cite{(Decadal12)}. A major quest, at the heart of many fascinating questions, will be to explain the
neutron-rich matter in the laboratory and the cosmos across a wide range of nucleonic densities. To this end, an interdisciplinary approach is essential  to integrate laboratory experiments with astronomical observations,  theory, and computational science.

In heavy neutron-rich nuclei, the excess of neutrons gives rise to 
a neutron skin, characterized by the neutron distribution extending beyond the proton  distribution. The skin can be characterized by
its thickness, which is commonly defined in terms of
the difference of neutron and proton point root-mean-square (rms) radii:
$\nskineq = \langle r_{\rm n}^{2} \rangle^{1/2}-\langle r_{\rm p}^{2}\rangle^{1/2}$. 
(As discussed in Ref.~\cite{(Miz00)}, it is better to define the neutron skin 
through neutron and proton diffraction radii and surface thickness. However, for well-bound nuclei, which do not exhibit halo features, the above definition of $r_\mathrm{skin}$ is
practically equivalent, see also \cite{(Rot09)}.)
The neutron-skin thickness
has been found to correlate with a number of
 observables in finite nuclei related to isovector nuclear fields \cite{(Bro00),(Fur02),(Rei10),(Roc11),(Pie12),(Fat13),(Naz13)}.
Furthermore, it has a close connection to the neutron matter equation of state (EOS)  and properties of  neutron stars 
\cite{(Ton84),(Rei99),(Hor01),*(Hor01a),(Typ01),(Fur02),(Yos04),(Sam09),(War09),(Rei10),(Fat12),*(Fat12a),(Agr12),(Lat12),(Ste13),(Erl13)}. 
In this context, precise experimental data on {\nskin} are indispensable;
they are crucial for  constraining the poorly known isovector  sector of nuclear structure models. 

Various experimental probes have been used to determine {\nskin}
\cite{(Dob96),(Miz00),(Pie12)}.
The PREX experiment has recently measured the parity-violating 
asymmetry coefficient $A_{\rm PV}$ for $^{208}$Pb \cite{(Abr12)}, which yielded $r_{\rm skin}$=0.33$^{+0.16}_{-0.18}$ \cite{(Hor12)}.

Unfortunately, the experimental error bar of PREX is too large to provide any practical constraint on well-calibrated theoretical models
\cite{(Pie12)}. At present, the most precisely determined \cite{(Tam11)} isovector indicator in heavy nuclei is the electric dipole polarizability $\alpha_{\rm D}$ in $^{208}$Pb \cite{(Rei10),(Rei13)}, which has been used to put constraints on  {\nskin} of $^{208}$Pb \cite{(Tam11),(Pie12)}. However, a number of important measurements are in the works.
A follow-up measurement to PREX, PREX-II \cite{Prex2}, has been 
designed to improve  experimental precision to 0.06\,fm. A CREX measurement of the neutron skin in $^{48}$Ca \cite{Crex} is promising an unprecedented precision of 0.02\,fm. Last but not least, on-going experimental studies of $\alpha_{\rm D}$ in several neutron-rich nuclei \cite{(Tam13)} will soon  provide key data.

The goal of this study is to survey {\nskin} across the nuclear landscape using  nuclear  Density Functional Theory (DFT)  \cite{(Ben03)} - a global theoretical approach
to  nuclear properties and  a tool of choice in microscopic studies of complex 
heavy nuclei. By considering several effective interactions, represented by different Skyrme energy density functionals (EDFs) optimized to experimental data, we assess the model (systematic) error on {\nskin}. Moreover, by means of the statistical covariance technique, we quantify statistical uncertainties of model predictions and identify those nuclear matter properties (NMP) of EDFs 
that are the main sources of statistical error.
In this way, we provide a benchmark for the precision of  future experiments on {\nskin} aiming at informing theory about isovector properties of effective nuclear interactions or functionals. 
This work builds on the previous global survey \cite{(Erl12)}, which
investigated model 
uncertainties on  drip-line positions and several global nuclear properties.
In particular, for the positions of the drip-lines, systematic and statistical errors were found to be quite similar giving us some confidence in the robustness of our extrapolations  into the {\it terra incognita} \cite{(Erl12),(Ols13)}. Here, we investigate whether the same also holds for  {\nskin}.


{\it Theoretical background} ---
The theoretical framework applied in this study is the same as in 
Refs.~\cite{(Erl12),(Ols13)}. Namely, we use self-consistent
 Hartree-Fock-Bogoliubov (HFB) theory with  six effective Skyrme interaction parameterizations  in the particle-hole channel
(SkM$^*$ \cite{(Bar82)}, SkP \cite{(Dob84)}, SLy4 \cite{(Cha98)}, SV-min \cite{(Klu09)}, {\UNEDFZERO} \cite{(Kor10)}, and {\UNEDFONE} \cite{(Kor12)})
augmented by the density-dependent, zero range
pairing interaction.
This set of EDF parameterizations has characteristics that are distinct enough  to  assess the systematic error within this family of Skyrme models.
The rms proton and neutron radii of even-even nuclei across the mass table were obtained in large-scale deformed  HFB calculations \cite{(Erl12a)}  using the solver {\HFBTHO} \cite{(Sto05)}. 
To approximately restore  the
particle number symmetry broken in HFB, we used the 
Lipkin-Nogami scheme of  Ref.~\cite{(Sto03)}. All remaining details are exactly as in Refs.~\cite{(Erl12),(Erl12a)}.

The Skyrme energy density is parameterized 
by about a dozen coupling constants that are determined by confronting DFT predictions with experiment. To relate to the nuclear matter EOS,  the volume part of the 
energy density is often  parameterized in terms of NMP \cite{(Klu09),(Kor10)}. 
Typically, the phenomenological  input used in parameter adjustment  consists of
nuclear masses and their differences, radii, surface thickness, mean energies of giant resonances, and other data (see Refs.~\cite{(Ben03),(Kor10),(Erl11)} 
for a list of observables commonly used in  the  EDF optimization).
The actual fit  is done by minimizing the objective function  
\begin{equation}\label{chi2}
 \chi^2(\mathbf{x})
=
\sum_p
\left(
\frac{\mathcal{O}_p^\mathrm{(th)}(\mathbf{x})
      -
      \mathcal{O}_p^\mathrm{(exp)}}
     {w_p}
\right)^2,
 \end{equation} 
with respect to EDF parameters $\mathbf{x}=\{x_i\}$. In Eq.~(\ref{chi2}),
${\cal O}_p$ is a selected  observable  and  $w_p$ is  
the corresponding weight that represents the adopted error.

Once the minimum 
$\mathbf{x}_{\rm min}$ of (\ref{chi2})
is found, a statistical covariance analysis can be carried out to obtain standard deviations and correlations 
between EDF parameters \cite{(Klu09),(Kor10),(Kor12),(Fat11)}.
The statistical standard deviation of an observable ${\cal O}$ is given by
\begin{equation}
\sigma_{\cal O}^{2} = \sum_{i,j} {\rm Cov}(x_{i},x_{j}) \left[\frac{\partial {\cal O}}{\partial x_{i}} 
                 \frac{\partial {\cal O}}{\partial x_{j}}\right],
\label{eq:stddev}
\end{equation}
where ${\rm Cov}(x_{i},x_{j})$ is the covariance matrix for the  model parameters. 
In the calculation of the covariance matrix, a linearized least-square system in the vicinity 
of the minimum $\mathbf{x}_{\rm min}$ is usually assumed. 
Within this approximation \cite{(Bra99)}
the covariance matrix is obtained in terms of the weights $w_p$ and 
the  partial derivatives $\partial_{x_i}\mathcal{O}_p^\mathrm{(th)}|_{\mathbf{x}_0}$, which 
are usually approximated by finite  differences.
Thus, the 
magnitude of the covariance matrix, and consequently the magnitude of the standard deviation
(\ref{eq:stddev}), depends on the chosen weights $w_{p}$.
The covariance matrix can be linked to the covariance ellipsoid between 
two parameters \cite{(Rei10)}.

Since the accuracy of  Skyrme EDFs, for example for nuclear binding energies, is usually worse 
compared to the  experimental error bars the weights $w_{p}$ should  be chosen to reflect
the expected accuracy of the model as opposed to the actual experimental error. As argued in Ref.~\cite{(Gao13)},
a balanced parameter optimization should lead to uncertainty comparable to the magnitude
of the optimization residuals. 
For example, with {\UNEDFZERO}, the residuals in binding energies are typically similar in magnitude
 to the adopted weights \cite{(Kor10)}.
In the optimization of SV-min, the adopted errors were  additionally scaled to give 
a lower weight to nuclei influenced by collective correlations \cite{(Klu09)}. With a proper choice of  weights, calculated statistical errors
(\ref{eq:stddev})  provide a realistic 
picture of theoretical  uncertainties and predictive capabilities of a model.

\begin{figure}[ht]
\center
\includegraphics[width=\linewidth]{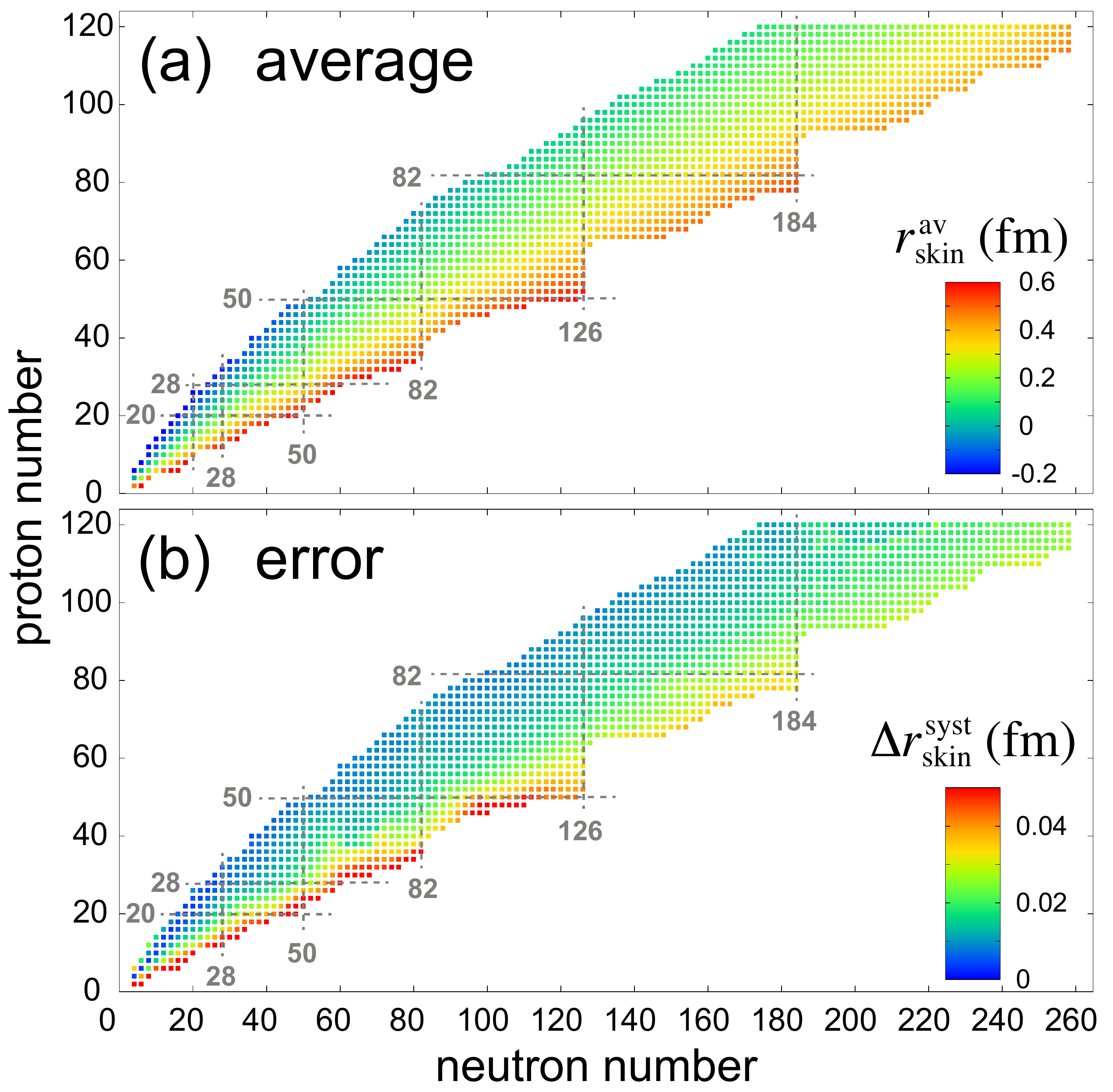}
\caption{(Color online)  (a) The model-averaged  value of {\nskin} and (b)
the systematic error $\Delta r_{\rm skin}^{\rm syst}$
for the six  EDFs used for the particle-bound even-even nuclei with $Z \le 120$.}
\label{fig:systematic}
\end{figure}

In the present work,  two different model uncertainties are considered.
The systematic  error  represents the rms spread of predictions of different Skyrme EDFs obtained by means of diverse fitting protocols. In the absence of the exact reference model, such an inter-model 
deviation represents a rough approximation to the systematic error, and it should be viewed as such. 
The statistical error represents the theoretical uncertainty associated with   model parameters and  
is obtained using least-squares covariance analysis \cite{(Klu09),(Rei10),(Rei13),(Gao13)}.



{\it Results} --- The mean value  of  {\nskin} and the corresponding rms  deviation  $\Delta r_{\rm skin}^{\rm syst}$ are shown in Fig.~\ref{fig:systematic} for all  even-even nuclei with $Z \le 120$ predicted to be particle-bound in all our models.
(The results for {\nskin} for 
each individual model can be found in the Supplementary Information of Ref.~\cite{(Erl12)}.)  As expected, 
the average value of the neutron skin thickness $r^{\rm av}_{\rm skin}$ increases steadily  with 
$N$  for each isotopic chain \cite{(Dob96),(Miz00)}. 
The systematic error  also increases gradually when approaching the neutron drip line. However, the range of $\Delta r_{\rm skin}^{\rm syst}$ 
is surprisingly small: the model spread does not exceed 0.05\,fm for extremely neutron-rich systems. This suggests that in spite of different optimization strategies, the EDFs considered give a very consistent answer when it comes to {\nskin}.

\begin{figure}[tbh]
\center
\includegraphics[width=\linewidth]{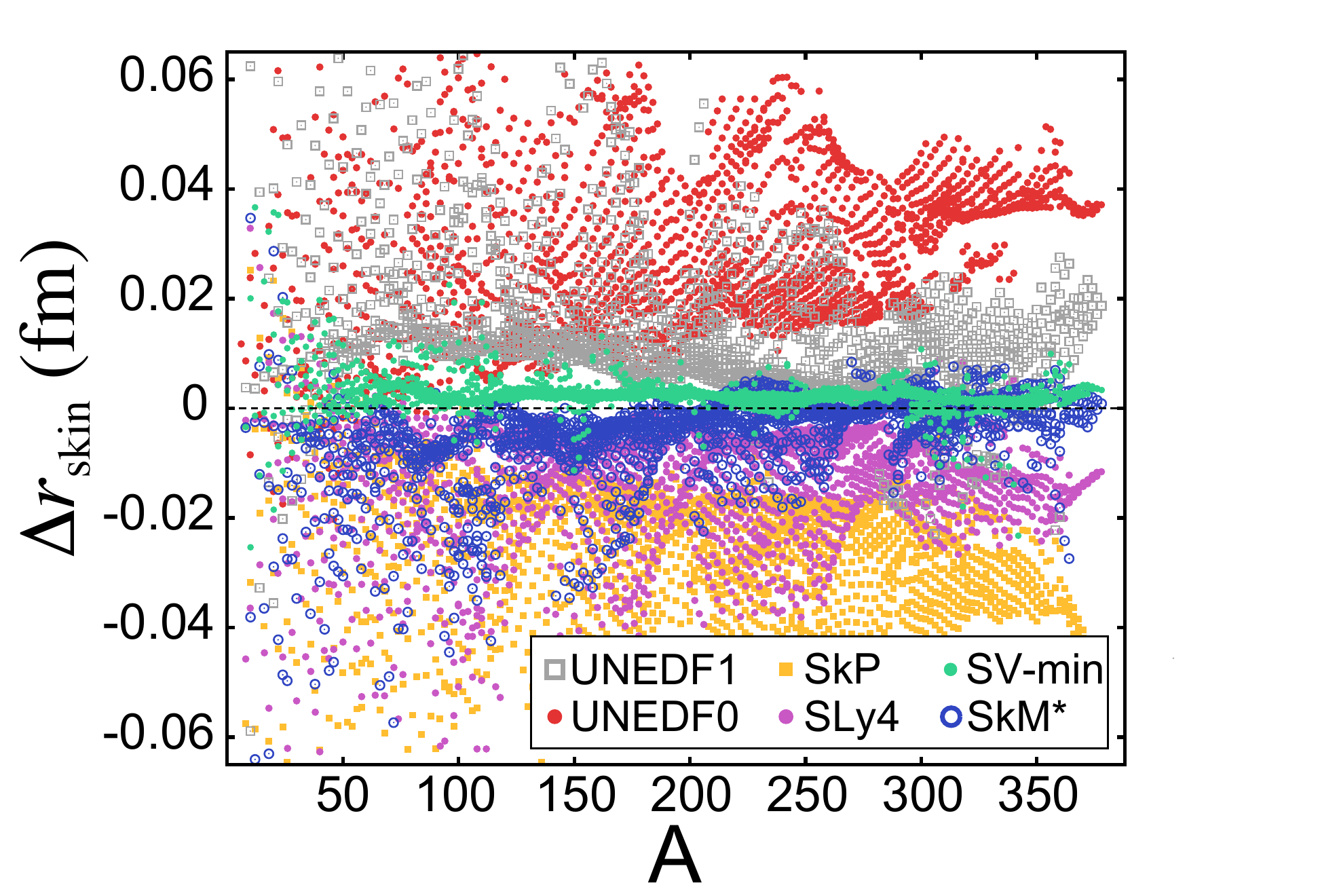}
\caption{(Color online) Scatter plot of the deviation of neutron skin thickness
from the mean value $r^{\rm av}_{\rm skin}$  for the six models used  as 
a function of mass number $A$.}
\label{fig:scatter}
\end{figure}
To get a deeper insight into the budget of $\Delta r_{\rm skin}^{\rm syst}$,  Fig.~\ref{fig:scatter} shows the individual residuals of {\nskin} with respect 
to $r^{\rm av}_{\rm skin}$. 
While SV-min  closely follows the  average trend,
SkP and {\UNEDFZERO} show large deviations.
By inspecting NMP of the  used EDFs \cite{(Rei06),(Kor10),(Kor12),(Erl13)}
one can see that low values of {\nskin} for SkP can be attributed to its particularly low value of the slope of the symmetry energy, $L=19.7$\,MeV (as compared to $L=44.8$\,MeV for SV-min). Still, the parameter $L$ cannot be the whole story, as - for instance - its value for {\UNEDFZERO}, $L=45.1$\,MeV is very close to that in SV-min.

\begin{figure*}[tbh]
\center
\includegraphics[width=\linewidth]{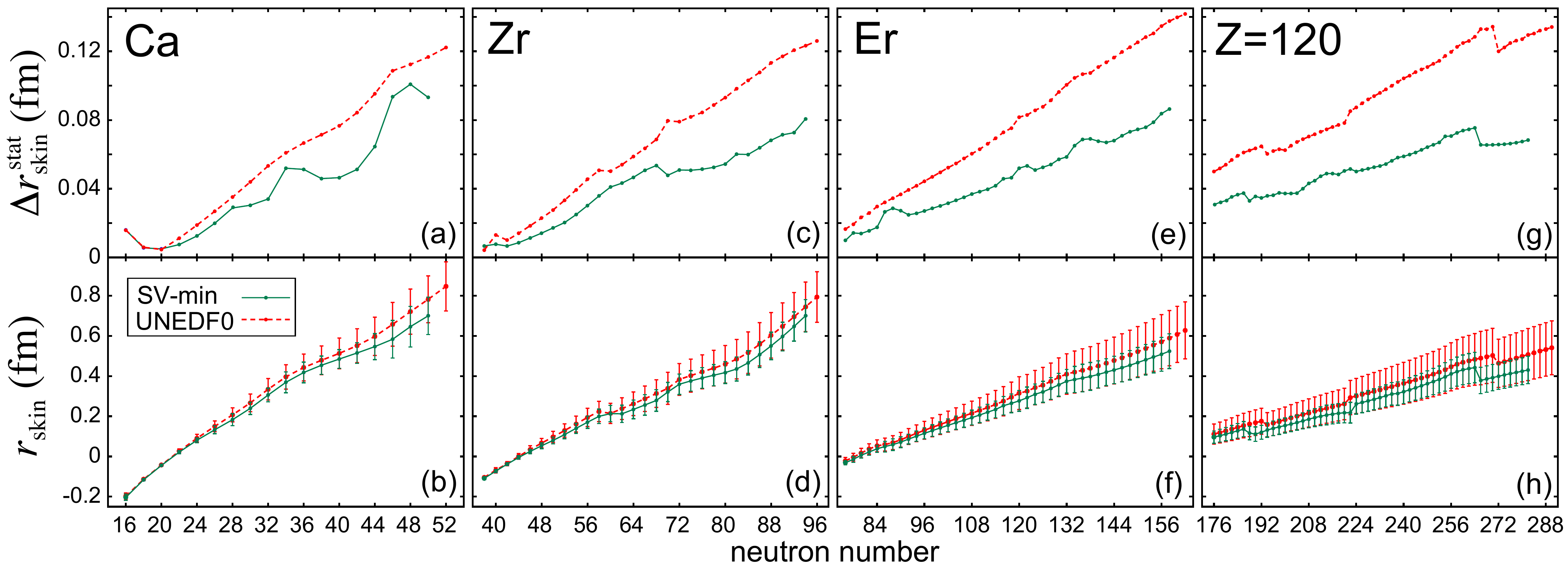}
\caption{(Color online) Top: calculated statistical error on {\nskin}
in Ca, Zr, Er, and $Z=120$ isotopic chains for {\UNEDFZERO} (dashed line) and SV-min (solid line).
Bottom:  corresponding  neutron skins with statistical uncertainties (error bars).}
\label{fig:statistic}
\end{figure*}

Figure~\ref{fig:statistic} shows the statistical error of {\nskin} 
for the isotopic chains of Ca, Zr, Er, and $Z=120$ obtained with  
{\UNEDFZERO} and SV-min. 
Even though the magnitude of statistical error, $\Delta r_{\rm skin}^{\rm stat}$,  is somewhat different for the  two models,
especially in the heavier isotopes, the model predictions for {\nskin} are consistent. 
Apparent discontinuities, e.g., for the $Z=120$ isotopic chain, are due to sudden changes 
in quadrupole deformation (see Ref.~\cite{(Erl12)}; supplementary information).
Also, similarly to the systematic error of Fig. \ref{fig:systematic}, $\Delta r_{\rm skin}^{\rm stat}$ propagates  with $N$. The gradual
growth of statistical error with the neutron excess is
primarily caused by the isovector coupling constants of the functional
that are poorly constrained by the current data.

The statistical error of {\UNEDFZERO} and SV-min on {\nskin} is significantly  larger as compared to the 
systematic error of Fig. \ref{fig:systematic}.
As discussed earlier, the statistical error of a computed  observable depends 
on the adopted errors used in  (\ref{chi2}). Since the  weights $w_p$
reflect the expected accuracy of the model, the error bars given in 
Fig.~\ref{fig:statistic} do provide a good measure of the model uncertainty.
The reason for the difference in the magnitude of $\Delta r_{\rm skin}^{\rm stat}$ between {\UNEDFZERO} and SV-min can be traced back
to the different optimization protocols in both cases.
Namely,  in the optimization of SV-min lower weights were 
assumed for certain nuclei to  account for collective correlations, and this explicitly impacts 
the standard deviation (\ref{eq:stddev}).
At the same time, the experimental data pool for SV-min includes, in addition to charge radii, 
diffraction radii and surface thickness  \cite{(Klu09)}, thus  reducing the 
statistical uncertainty compared to {\UNEDFZERO}.

\begin{figure}[tbh]
\center
\includegraphics[width=\linewidth]{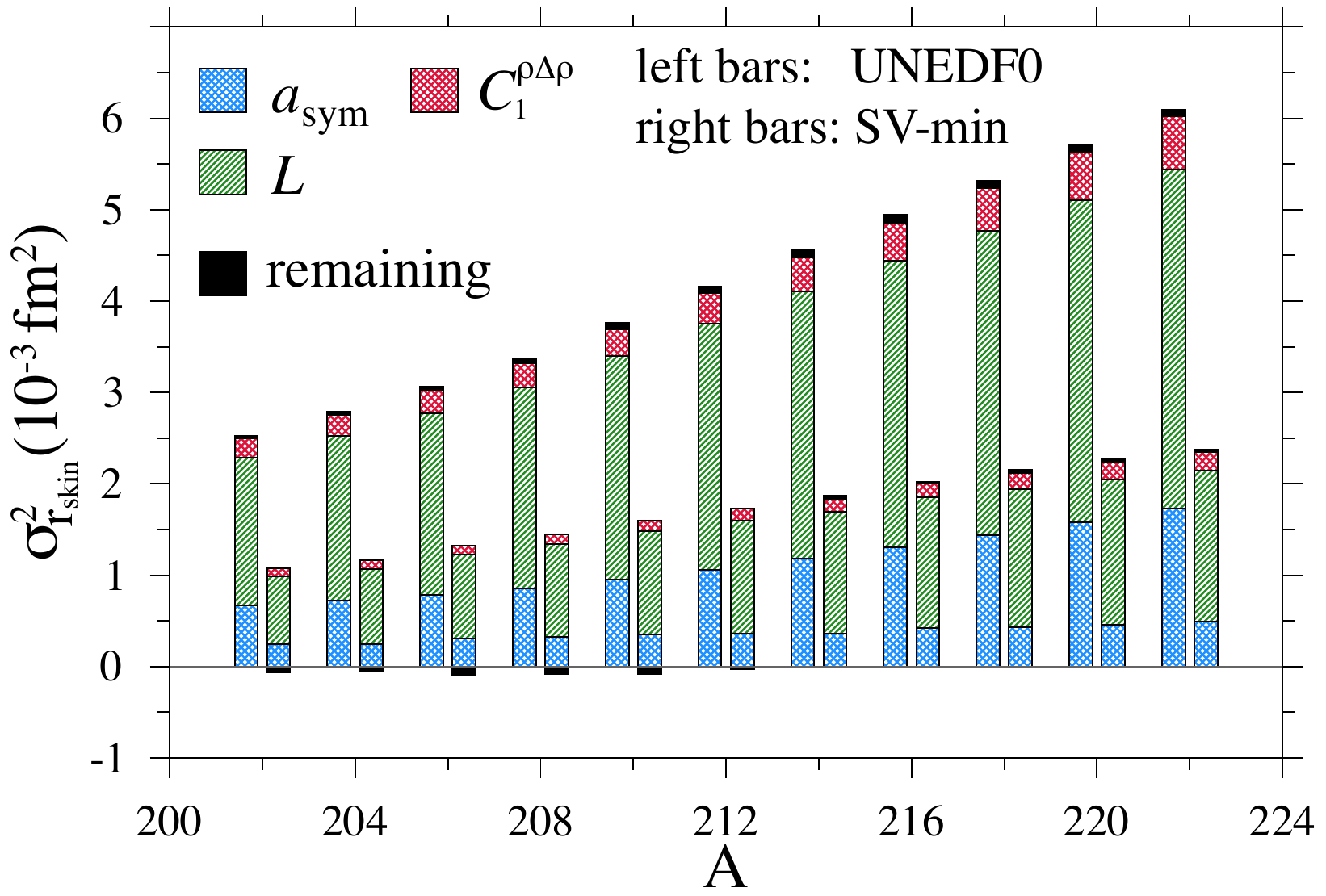}
\caption{(Color online) Error budget of $\sigma^2_{r_{\rm skin}}$
(\ref{eq:stddev})
for {\UNEDFZERO} (left bars)  and SV-min (right bars)  for even-even isotopes of Pb.
The dominant sources of uncertainty are the symmetry energy parameter $a_{\rm sym}$, the slope of the symmetry energy $L$, and -- to a lesser degree -- the isovector coupling
constant $C^{\rho\Delta \rho}$ that governs the surface properties of EDFs.}
\label{fig:contribution}
\end{figure}

For {\UNEDFZERO} and SV-min, the dominant contributions to $\Delta r_{\rm skin}^{\rm stat}$ come from  $L$ and $\asym$. 
This is illustrated
 in Fig.~\ref{fig:contribution}, where the contributions
to the sum of Eq. (\ref{eq:stddev}) are plotted for Pb isotopes
(the second index
$j$ is summed over all the parameters). The contribution from $L$ is by far the largest one
in all isotopes, and it yields over 50\%  of the total error.
We checked that this  also holds for other semi-magic isotopic chains.
The strong impact of $L$ on the statistical
error of neutron rms radii was also found in Ref.~\cite{(Gao13)}.
For SV-min, some of the isoscalar coupling constants also provide contributions comparable in the
magnitude to the $\asym$ parameter. However, when these contributions are summed up, they cancel out rather precisely and the net value is small. This is expected, since
 correlations between isoscalar and isovector parameters in SV-min are low
 \cite{(Rei10)}. 
 
While $\asym$ is determined fairly precisely for both {\UNEDFZERO}
($30.5 \pm 3.1$\,MeV) and SV-min ($30.7 \pm 1.9$\,MeV), the uncertainty in $L$ is much greater: $L=45 \pm 40$\,MeV  and $45 \pm 26$\,MeV for {\UNEDFZERO}
and SV-min, respectively. The fact that the symmetry energy and its slope are less precisely determined in {\UNEDFZERO} is reflected in the larger error  $\Delta r_{\rm skin}^{\rm stat}$ 
seen in   Fig.~\ref{fig:contribution}.

\begin{table}[bth]
\begin{center}
\caption{Theoretical uncertainties on {\nskin} in $^{208}$Pb and $^{48}$Ca (in 
fm). Shown  are statistical errors of {\UNEDFZERO} and SV-min, systematic 
error $\Delta r_{\rm skin}^{\rm syst}$,  the model-averaged deviation of Ref.~\cite{(Pie12)}, and  errors of PREX  \cite{(Abr12)} and planned PREX-II
\cite{Prex2} and CREX \cite{Crex} experiments.}
\label{table:uncert}
\begin{ruledtabular}
\begin{tabular}{lccccc}
 \multirow{2}{*}{nucleus} & \multicolumn{2}{c} {$\Delta r_{\rm skin}^{\rm stat}$}   &  \multirow{2}{*}{$\Delta r_{\rm skin}^{\rm syst}$}  &  \multirow{2}{*}{Ref.~\cite{(Pie12)}} & \multirow{2}{*}{Experiment}\\
                 & {\UNEDFZERO} & SV-min  &   & \\
\hline
 &  & & & \\[-8pt]
$^{208}$Pb & 0.058 & 0.037 & 0.013 & 0.022 & 0.18 \cite{(Abr12)}, 0.06\cite{Prex2}  \\
$^{48}$Ca  & 0.035 & 0.026 & 0.019 & 0.018 & 0.02 \cite{Crex} \\
\end{tabular}
\end{ruledtabular}
\end{center}
\end{table}

Finally, to address the required experimental accuracy to constrain Skyrme EDF models by future measurements of {\nskin}, we present in Table~\ref{table:uncert} $\Delta r_{\rm skin}^{\rm stat}$ of {\UNEDFZERO} and SV-min, the 
systematic error $\Delta r_{\rm skin}$, and  the model-averaged deviation of  Ref.~\cite{(Pie12)} constrained by the measured value of $\alpha_{\rm D}$ in $^{208}$Pb \cite{(Tam11)}.
The results are presented for $^{208}$Pb and  $^{48}$Ca. 
The error bar of PREX \cite{(Abr12)} is unfortunately too large ($\sim$0.18\,fm) to provide a useful constraint on isovector properties of current models.
On the other hand, the superb anticipated accuracy of the planned CREX experiment (0.02\,fm) \cite{Crex} will have an impact on reducing the statistical error
on {\nskin}. 


{\it Conclusions} ---
This survey addresses  systematic and statistic errors on the 
neutron skin thickness predicted by various Skyrme EDF models. Because {\nskin} 
has been found to strongly correlate  with various isovector indicators,
it  provides an essential  constraint on nuclear EDFs that aim at making 
extrapolations into the  {\it terra incognita} at the neutron-rich side of the nuclear landscape.
We have found that systematic  error $\Delta r_{\rm skin}^{\rm syst}$ obtained in this work and in Ref.~\cite{(Pie12)} 
is  smaller than  the statistical error $\Delta r_{\rm skin}^{\rm stat}$. 
As expected, both errors grow with neutron number due to propagation
of uncertainties of poorly determined EDF isovector coupling constants.

The slope of the symmetry energy $L$ is the single main contributor to $\Delta r_{\rm skin}^{\rm stat}$.  As already pointed out in many
previous studies, this parameter is strongly correlated with many isovector indicators. Therefore, planned precise measurements of {\nskin} will help in pinning down this crucial NMP.
Conversely, if $L$ could be constrained by some other experimental data
\cite{(Ste13)}, 
this would also reduce  model uncertainty on {\nskin}. The methodology presented in this paper aiming at assessing statistical and systematic uncertainties on calculated quantities can be generally used to determine  the uniqueness and usefulness of an observable with respect to current theoretical models and can be used to help in planning future experiments and experimental
programs \cite{(Maj13)}.


\begin{acknowledgments}
This work was supported by the U.S. Department of Energy (DOE) under Contracts No.
DE-FG02-96ER40963 (University of Tennessee) and No. DE-SC0008499    (NUCLEI SciDAC Collaboration),
by the Academy of Finland under the Centre of Excellence Programme
2012--2017 (Nuclear and Accelerator Based Physics Programme at JYFL) and 
FIDIPRO programme,
and by the European Union's Seventh Framework Programme ENSAR (THEXO) 
under Grant No. 262010.
Computational resources were provided through an INCITE award ``Computational
Nuclear Structure" by the National Center for Computational Sciences (NCCS) 
and the National Institute for Computational Sciences (NICS) at Oak Ridge 
National Laboratory.

\end{acknowledgments}

\bibliographystyle{apsrev4-1}

\end{document}